\documentclass[12pt]{article}
\usepackage{amsmath,amssymb}
\newcommand{\be}{\begin{equation}}
\newcommand{\bea}{\begin{eqnarray}}
\newcommand{\ena}{\end{eqnarray}}

\newcommand{\e}{{\rm e}}

\newcommand{\M}{{\cal M}}

\newcommand{\bx}{{\bf x}}
\newcommand{\bp}{{\bf p}}
\newcommand{\der}{\partial}

\newcommand{\RR}{\mbox{${\mathbb R}$}}

\begin{document}

\thispagestyle{empty}

\rightline {IFUP-TH 34/2001}

\vskip 1 truecm
\centerline {\Large \bf The Role of Initial Conditions}
\bigskip
\centerline{\Large \bf in Presence of Extra Dimensions}
\vskip 1.5 truecm
\centerline {\large \rm Mihail Mintchev}
\medskip
\bigskip
\centerline {\it Istituto Nazionale di Fisica Nucleare, Sezione di Pisa}
\centerline {\it Dipartimento di Fisica dell'Universit\`a di Pisa,}
\centerline {\it Via Buonarroti 2, 56127 Pisa, Italy}
\bigskip
\medskip

\vskip 3.5 truecm
\centerline {\large \it Abstract}
\medskip
\bigskip
Quantum field theory in 4+1 dimensional bulk space with boundary,
representing a $3$-brane, is considered. We study the impact of the initial
conditions in the bulk on the field dynamics on the brane.
We demonstrate that these conditions determine the Kaluza-Klein measure.
We also establish the existence of a rich family of quantum fields on 
the brane,
generated by the same bulk action, but corresponding to different 
initial conditions.
A simple classification of these fields is proposed and it is shown
that some of them lead to ultraviolet finite theories, which have some
common features with strings.
 
\bigskip

\bigskip
\bigskip
\bigskip

\centerline {October 2001}
\vfill \eject

\section{\bf Introduction}

\bigskip

There have been in the past various attempts to render the short 
distance properties of
quantum fields milder by modifying or relaxing the requirement of 
local commutativity.
The first efforts in this direction can be traced back to the
late forties \cite{Snyder:1947qz, Y} and early fifties \cite{Pa,CP}, when
such relevant ideas like quantized space-time,
nonlocal formfactors and the concept of weak local commutativity 
\cite{Dyson:1958gx}
already emerged. The modern approach to the subject
is dominated by string theory and the deeply related field theory on
noncommutative space-time.
In this paper we discuss locality in the light
of the recent developments in quantum field
theory with extra dimensions and branes
\cite{Arkani-Hamed:1998rs} - \cite{Anchordoqui:2000du},
focusing mainly on the role of the initial conditions.

In order to fix the ideas, we consider a toy model in 
five-dimensional bulk space
$\M = \RR^{4} \times \RR_+$, where $\RR_+$ is the half line $\{y\in 
\RR \, :\, y>0 \}$.
We adopt  the coordinates
$(x,y) \in \RR^{4} \times \RR_+$ with $x \equiv (x^0,x^1,x^2,x^3) = (x^0,\bx)$
and a diagonal flat metric $G_{\alpha \beta }$ ($\alpha , \beta = 
0,...,4$) with
${\rm diag}\, G = (1,-1,-1,-1,-1)$.
The boundary $\der M$ coincides with the $3+1$-dimensional Minkowski space
${\bf M}_{3+1}$, modeling a static $3$-brane. The
dynamics of the model is defined by the action
\be
S = \frac{1}{2} \int_{-\infty}^\infty d^4x  \int_0^\infty dy
\left ( \der^\alpha \Phi \der_\alpha \Phi - M^2 \Phi^2\right ) (x,y)
  - \int_{-\infty}^\infty d^4x \left(\frac{\eta }{2} \Phi^2
- \frac{g}{3!}\Phi^3 \right) (x,0) \, ,
\label{act}
\end{equation}
whose variation gives the equation of motion
\be
\left (\der^\alpha \der_\alpha + M^2 \right ) \Phi (x,y) =  0 \, , 
\qquad y>0\, ,
\label{eqm}
\end{equation}
and the boundary condition
\be
\left [\left (\der_y - \eta \right )\Phi +  \frac{g}{2} \Phi^2 
\right] (x,0) = 0 \, .
\label{bcphi}
\end{equation}
Being localized on the boundary, the interaction generates a nonlinear
boundary condition. We will work out explicitly the
case $\eta \geq 0$, commenting at the end of the paper on the range $\eta < 0$.
The action (\ref{act}) breaks down all translations, rotations and
Lorentz transformations involving the $y$-coordinate. Notice, however, that
brane Poincar\'e invariance is preserved.

Eqs.(\ref{eqm},\ref{bcphi}) define an initial boundary value problem.
Accordingly, the solution depends on the initial conditions fixed
on a time-like hypersurface in $\M$. In the context of quantum field theory
these conditions are determined by the equal-time commutation relations.
The general form of the nontrivial commutator in our case is
\bea
\left [(\der_0\Phi )(x^0,\bx_1,y_1)\, ,\, \Phi (x^0,\bx_2,y_2)\right ] =
-i\delta (\bx_1-\bx_2)\, \xi (y_1,y_2) \, , \qquad y_1,\, y_2 > 0 \, .
\label{ccr}
\ena
The distribution $\xi$ plays an essential role in what follows.
The basic physical requirements of positivity
and locality provide some restrictions on $\xi$, but do not
fix $\xi$ completely. The asymmetry of $\M$, caused by the extra 
dimension $y\in \RR_+$,
implies the existence of inequivalent classes of $\xi$-distributions,
inducing different ultraviolet (UV) behavior on the brane. Our main goal
below will be to establish and investigate this new phenomenon, which
has no counterpart in local relativistic Lagrangian field theory on 
${\bf M}_{3+1}$.
We will show that the dynamics on the brane ${\bf M}_{3+1}$ strongly
depends on the initial conditions in the bulk $\M$.

Let us analyze first the impact of positivity on $\xi$. Treating the 
interaction
perturbatively in $g$, we start with the case $g=0$. We first introduce
the complete orthonormal set of eigenfunctions ($\lambda >0$)
\be
\psi (y,\lambda)  = \e^{-i\lambda y} + B(\lambda )\, \e^{i\lambda y}\, ,
\qquad B(\lambda ) \equiv \frac{\lambda - i\eta}{\lambda + i\eta}\, ,
\label{system}
\end{equation}
of the operator $-\der_y^2$ on $\RR_+$, whose domain is defined
in agreement with (\ref{bcphi}) by the
boundary condition
\be
\lim_{y \downarrow 0} \left (\der_y - \eta \right )\psi (y)  = 0 \, .
\label{bc}
\end{equation}
Then $\xi (y_1,y_2)$ is generated by a distribution $\sigma $ of
the single variable $\lambda \in \RR_+$ via
\be
\xi (y_1,y_2) = \int_0^\infty \frac{d\lambda}{2\pi}\, {\overline 
\psi}(y_1,\lambda )
\sigma (\lambda ) \psi (y_2,\lambda ) \, .
\label{xi}
\end{equation}
The requirement of positivity of the metric in the state space takes 
a simple form
in terms of $\sigma $ and reads
\be
\sigma (\lambda )\geq 0
\label{pos}
\end{equation}
on $\RR_+$. We assume (\ref{pos}) in what follows and proceed by 
clarifying the physical
content of $\sigma$. It is convenient for this purpose to compute
the two-point vacuum expectation value (Wightman function) of
$\Phi $. The result is \cite{Mintchev:2001yz}
\bea
\langle \Phi (x_1,y_1) \Phi(x_2,y_2) \rangle_{{}_0} =
\int_0^\infty \frac{d\lambda}{2\pi}\, {\overline \psi}(y_1,\lambda )
\sigma (\lambda ) \psi (y_2,\lambda )\, W_{M^2+\lambda^2}(x_{12}) \, ,
\label{ppb}
\ena
where
\be
W_{m^2}(x)
= \int_{-\infty}^\infty \frac{d^4p}{(2\pi)^4} \,
\e^{-ipx} \theta (p^0) 2\pi \delta (p^2 - m^2) \,
\label{w}
\end{equation}
is the familiar two-point scalar function of mass $m^2$ in ${\bf M}_{3+1}$.
It is perhaps useful to recall that
the integral over $\lambda $ in (\ref{ppb}) and in what follows
must be interpreted in the sense of
distributions, i. e. after smearing the integrand with suitable test 
functions over
$\M$. The choice of the test function space depends on the behavior of
$\sigma $ for $\lambda \to \infty $ and will be specified below.

Eq.(\ref{ppb}) completely fixes the correlation functions of $\Phi$ for $g=0$.
The latter determine the quantum field $\varphi $, induced on the 
brane, by means of
\be
\langle \varphi (x_1)\cdots \varphi (x_{n})
\rangle_{{}_0} =
\lim_{y_i \downarrow 0}\, \langle \Phi (x_1,y_1)\cdots \Phi 
(x_{n},y_{n}) \rangle_{{}_0} \, .
\label{corg}
\end{equation}
The limit (\ref{corg}) defines integral representations for the 
correlation functions of
$\varphi$, whose fundamental features are encoded in the two-point function
\be
w(x_{12}) = \langle \varphi (x_1)\varphi (x_2)\rangle_{{}_0} =
\int_0^\infty d\lambda^2 \, \varrho (\lambda^2 )\, W_{\lambda^2}(x_{12}) \, ,
\qquad x_{12}\equiv x_1-x_2 \, ,
\label{bpp0}
\end{equation}
with
\be
\varrho (\lambda^2 ) = \theta (\lambda^2 - M^2)
\frac{\sqrt {\lambda^2-M^2}\, \sigma (\sqrt {\lambda^2 -M^2}\, )}
{\pi(\lambda^2+\eta^2-M^2)} \, .
\label{ro}
\end{equation}
We observe in passing that the expansion of $\varrho $ around
$\eta = 0$ produces spurious divergences at $\lambda^2 = M^2$ in 
(\ref{bpp0}). This
feature represents a serious obstruction to the perturbative 
treatment \cite{Goldberger:2001tn} of
the boundary term $\frac{\eta}{2} \Phi^2$ in the action (\ref{act}). 
Using (\ref{w}),
the integral in (\ref{bpp0}) can be easily performed in momentum space and for
the Fourier transform ${\widehat w}$ of $w$ one finds
\be
{\widehat w}(p) = 2\theta (p^0)
\frac{ \theta (p^2-M^2) \sqrt {p^2-M^2}}
{p^2-M^2 + \eta^2}\, \sigma (\sqrt {p^2 -M^2}\, ) \, .
\label{ft1}
\end{equation}
Eqs.(\ref{bpp0}-\ref{ft1}) establish the physical meaning of $\sigma 
$. We deduce that the
mass spectrum of $\varphi$ belongs to $[M^2,\, \infty)$. The corresponding
eigenstates represent the Kaluza-Klein (KK) modes, $\sigma $ defining 
the weight
each KK mode contributes to the field $\varphi $.
The initial conditions in the bulk determine therefore the KK measure 
$\varrho $
on the brane. This result provides valuable information about a
central problem in quantum field theory with extra dimensions - the integration
over the KK modes. In fact, the KK measure is an essential element for the
complete understanding of this recently debated problem.

In order to classify the possible initial conditions (KK measures),
one can adopt the concept of local commutativity. Let us introduce 
for this purpose
some terminology. We will say that an initial condition
(\ref{ccr}) is of type A, if the corresponding bulk field $\Phi $ is 
local. Otherwise,
we call the initial condition of type B. One can further refine this 
classification
using the locality properties of $\varphi$. Since in passing from the 
bulk to the
brane one is suppressing a spatial dimension, type A initial conditions
generate only local brane fields. This is not the case for type B conditions.
In fact, there exist \cite{Mintchev:2001yz} nonlocal bulk fields 
which induce local brane fields.
We call the relative initial conditions of type ${\rm B}_1$. Finally, 
if both $\Phi $ and
$\varphi $ are nonlocal, one has type ${\rm B}_2$ initial conditions. We shall
illustrate below this elementary classification, focusing in each case on
the UV behavior of $\varphi$.

Consider first type A initial conditions. Since the
bulk hypersurface $x^0 = {\rm const.}$ is space-like, local 
commutativity of $\Phi $
applied to (\ref{ccr}) implies that the support of $\xi $ is in 
$y_1=y_2$. A large class
of such distributions is obtained by setting
\be
\sigma (\lambda ) = P(\lambda^2 ) \, ,
\label{p}
\end{equation}
where, according to (\ref{pos}), $P$ is any positive definite polynomial.
In fact, from (\ref{xi}) one gets
\be
\xi(y_1,y_2) = P(-\der^2_{y_1})\, \delta (y_{12}) \, ,
\label{xiloc}
\end{equation}
which has the required support property and generates
a local bulk field $\Phi $. This feature can be checked directly, taking
into account \cite{Liguori:1998xr, Liguori:1998vd} that the signals
propagating in the bulk are also reflected from the boundary, the reflection
coefficient being $B(\lambda )$.

To our knowledge only the case
\be
P(\lambda^2) = {\rm const.} > 0
\label{p1}
\end{equation}
has been explored till now in the literature. With this choice the propagator
\be
\tau(x) \equiv i\theta (x^0) w(x) + i \theta (-x^0) w(-x)
\label{tau}
\end{equation}
has the milder UV behavior within the set of initial
conditions defined by eqs.(\ref{ccr},\ref{xiloc}).
Indeed, using (\ref{ft1}) one gets
\be
{\widehat \tau } (p) =
  - \frac{\sqrt{M^2 - p^2} - \eta}{p^2 -M^2 + \eta^2 + i \varepsilon 
}\, P(p^2-M^2) \, .
\label{taup}
\end{equation}
Because of positivity, the optimal decay property of the right hand
side of (\ref{taup}) for large  Euclidean momenta $p_{{}_E} = (-ip^0, \bp)$
is obtained in the case (\ref{p1}). This is the only motivation we 
are aware for
selecting (\ref{p1}) among all positive definite polynomials.
Eq.(\ref{taup}) implies also that the boundary interaction ($g\not=0$) in
(\ref{act}) is nonrenormalizable for all initial conditions of the form
(\ref{ccr},\ref{xiloc}). Moreover, the relative brane fields
$\varphi $ are not canonical.

Concerning the type B initial conditions, we start with the simple example
\be
\sigma (\lambda ) = \lambda \, , \qquad \eta = 0 \, .
\label{II1}
\end{equation}
One easily derives (Pv stands for principal value)
\be
\xi (y_1,y_2) = -\frac{1}{\pi}\left ( {\rm Pv}\, \frac{1}{y_{12}^2} +
{\rm Pv}\, \frac{1}{{\widetilde y}_{12}^2} \right ) \, ,
\qquad {\widetilde y}_{12} \equiv y_1+y_2 \, ,
\label{II2}
\end{equation}
whose support is not localized at $y_1=y_2$. Therefore, the 
corresponding bulk field
$\Phi$ is nonlocal. In spite of this fact, eqs.(\ref{bpp0},\ref{ro}) give
\be
\langle \varphi (x_1)\varphi (x_2)\rangle_{{}_0} =
\frac{1}{\pi}\int_0^\infty d\lambda^2 \,  \theta (\lambda^2 - M^2)\, 
W_{\lambda^2}(x_{12}) \, ,
\label{II3}
\end{equation}
which defines a local generalized free field \cite{J} on the brane. 
Thus, the $y\to 0$ limit
absorbs all nonlocal effects present in the bulk and, according to 
our classification,
the initial condition (\ref{II1}) is of type ${\rm B}_1$. Such 
conditions have been
discovered and described in some detail in \cite{Mintchev:2001yz}.
A remarkable subset of the class ${\rm B}_1$ is obtained by requiring
\be
\int_0^\infty d\lambda^2 \, \varrho (\lambda^2 ) \equiv C < \infty \, .
\label{cbound}
\end{equation}
Besides of being local, the field $\varphi $ induced on the brane is 
in this case
also canonical, namely
\be
\left [(\der_0\varphi) (x^0,\bx_1)\, ,\, \varphi (x^0,\bx_2)\right ] =
-i\, C\delta (\bx_1-\bx_2)\, .
\label{ppcc}
\end{equation}
The bound (\ref{cbound}) implies also that the brane propagator satisfies
\be
p_{{}_E}^2\, {\widehat \tau } (p_{{}_E}) \leq C \, .
\label{est}
\end{equation}
Thus ${\widehat \tau } (p_{{}_E})$ decays at least like
$1/{{p_{{}_E}^2}}$ when $p_{{}_E}^2 \to \infty$,
leading to a renormalizable perturbative expansion for the
correlation functions of our model (\ref{act}) on the brane. It has been shown
in \cite{Mintchev:2001yz}, that this UV behavior cannot be improved 
further within the class ${\rm B}_1$.
We conclude therefore that relaxing local commutativity in the bulk, 
but keeping this
requirement on the brane, allows to generate at most renormalizable 
brane interactions.
Bulk fields of the type ${\rm B}_1$ can be constructed
\cite{Mintchev:2001yz} on an AdS background as well and
we expect that they may help for evading some no-go theorems
\cite{Maldacena:2001mw}, concerning
the construction of Randall-Sundrum compactifications from supergravity.

Conformal invariance on the brane deserves also a special comment.
Conformal covariant fields with anomalous
dimension $d>1$ are generated for $M = \eta = 0$ by initial 
conditions defined by
\be
\sigma (\lambda ) = \frac {\pi}{\Gamma (d-1)}\, \lambda^{2d-3} \, .
\label{conf}
\end{equation}
Except for $d=k + \frac{3}{2}$ with $k=1,2,...$, all these fields
are induced by nonlocal bulk fields, which confirms the relevance of
giving up local commutativity in the bulk.

A common feature of type A and type ${\rm B}_1$ initial conditions is 
that both of them can be
characterized by requiring $\sigma $ to be polynomially bounded.
As test functions in (\ref{ppb},\ref{bpp0}) one can use
smooth functions of rapid decrease belonging to the Schwartz test function
space $\cal S$ \cite{GSch}. Accordingly, the correlation functions are
tempered distributions.

We turn now to type ${\rm B}_2$ initial conditions. Generalizing 
eq.(\ref{xiloc}), we
consider the series
\be
\xi(y_1,y_2) = \sum_{n=0}^\infty \frac{c_n}{(2n)!}
(-\der^2_{y_1})^n \, \delta (y_1-y_2)\, ,
\label{xilocinf}
\end{equation}
with a restriction on the growth of the coefficients $c_n$ for $n\to \infty $.
Let us assume for instance that
\be
{\overline \lim}_{n\to \infty}|c_n|^{1/n} n^{-a} < \infty \, ,
\label{estE}
\end{equation}
{}for some $0<a<2$. Such kind of distributions have been already
applied in nonlocal quantum field theory by Efimov \cite{E,Eb}.
Since the corresponding $\sigma $ may grow exponentially for $\lambda 
\to \infty$, the
test functions in momentum space must be of compact support.
Consequently \cite{GSch}, the correlators in coordinate space define analytic
functionals instead of tempered distributions.
Referring for the details
about such functionals to the mathematical literature (see e.g. \cite{GSch}),
we prefer to concentrate below on the explicit example
\be
\sigma (\lambda ) = \e^{l^2 \lambda^2} \, ,
\label{ex}
\end{equation}
where $l\geq 0$ is a free parameter with dimension of length.
Notice that $l=0$ reproduces the conventional initial condition
$\xi(y_1,y_2)= \delta(y_{12})$. The requirement (\ref{estE}) is
satisfied for $1\leq a <2$ and inserting (\ref{ex}) in (\ref{ft1}),
one gets the two-point function
\be
{\widehat w}(p) = 2\theta (p^0)
\frac{ \theta (p^2-M^2) \sqrt {p^2-M^2}}
{p^2-M^2 + \eta^2}\, \e^{l^2 (p^2 -M^2)} \, ,
\label{ft2}
\end{equation}
carrying the complete information about $\varphi$ for $g=0$. As 
expected, both $\Phi$ and
$\varphi$ are nonlocal.

Switching on the interaction, one can analyze the model (\ref{act}) in
perturbation theory. From eqs.(\ref{tau},\ref{ft2}) one obtains the propagator
\be
{\widehat \tau } (p) =
  - \frac{\sqrt{M^2 - p^2} - \eta}{p^2 -M^2 + \eta^2 + i \varepsilon }\,
\e^{l^2(p^2 -M^2)}\, .
\label{taup1}
\end{equation}
We see that ${\widehat \tau}(p_{{}_E})$ decays exponentially for 
$p_{{}_E}^2 \to \infty$,
which implies that $\varphi $ has UV finite perturbative expansion. 
In order to get an idea
about this expansion, let us derive the one-loop self energy $\Sigma^{(1)}$.
In the case $\eta = 0$ one has
\be
\Sigma^{(1)} (p) = ig^2 \int \frac{d^4k}{(2\pi)^4}
\frac{\sqrt{(M^2-k^2)[M^2-(p-k)^2]}}
{(k^2 -M^2 + i \varepsilon )[(p-k)^2 -M^2 + i \varepsilon ]}\,
\e^{l^2[k^2 + (p-k)^2 -2M^2]}\, .
\label{self1}
\end{equation}
The integration in $k$ can be performed by means of the integral representation
\be
\frac{\sqrt D}{D+i \varepsilon} = \frac{1}{\sqrt {i\pi}}\int_0^\infty 
d\alpha \,
\frac{\e^{i\alpha (D+i\varepsilon)}}{\sqrt \alpha}\, , \qquad 
\varepsilon > 0,\quad D\in \RR \, .
\label{irep}
\end{equation}
One finds
\be
\Sigma^{(1)} (p) = \frac{ig^2}{16\pi^3} \int_0^\infty d\alpha_1 
\int_0^\infty d\alpha_2
\, \frac{\e^{-\varepsilon (\alpha_1+\alpha_2)}}
{\sqrt {\alpha_1 \alpha_2}}\, F(\alpha_1-il^2, \alpha_2-il^2) \, ,
\label{self2}
\end{equation}
with
\be
F(\alpha_1,\alpha_2) = \frac{\e^{ip^2\frac{\alpha_1\alpha_2}{\alpha_1+\alpha_2}
-iM^2(\alpha_1+\alpha_2)}}{(\alpha_1+\alpha_2)^2}\, .
\label{self3}
\end{equation}
The effect of the exponential factor in the propagator (\ref{taup1})
is the $il^2$ shift in both arguments
of $F$ in (\ref{self2}). Standard manipulations lead at this stage to
the estimate ($\varepsilon \to 0$)
\be
|\Sigma^{(1)} (p)| < \frac{g^2}{l^2}\, \e^{l^2(p^2-2M^2)} \, ,
\label{self4}
\end{equation}
which confirms that $\Sigma^{(1)}$ is finite for $l\not=0$.
The argument can be easily extended to the one-loop vertex function 
$\Lambda^{(1)}$
and applied to any order in perturbation theory. The key point is obviously
the exponential factor in the propagator (\ref{taup1}).
It is instructive to recall in this respect that a similar mechanism
works in string field theory as well. Indeed, the string vertices
involve \cite{Cremmer:1986if, Gross:1987ia} the factor $\exp 
{(-\alpha^\prime p_{{}_E}^2)}$,
causing loops to converge.
In spite of the fact that in our case such a factor enters the propagator,
the net result is the same: the perturbative expansion of (\ref{act}), with
initial conditions defined by (\ref{ccr},\ref{xi},\ref{ex}), is
UV finite order by order. This result shows that by an appropriate
choice of the initial conditions in the bulk,
one can reproduce some features of string theory in the context
of quantum field theory with extra dimensions. In this spirit it
is tempting to relate the parameter $l$ to the inverse of the string tension
$\alpha^\prime$ by $l=\sqrt{\alpha^\prime}$.

An interesting boundary phenomenon takes place \cite{Liguori:1998xr} 
in the range $\eta < 0$.
One can imagine heuristically, that the boundary develops a sort of 
attractive force,
producing a boundary bound state. Indeed, in addition to the scattering
states (\ref{system}), the operator $-\der_y^2$ admits for $\eta < 0$ 
the bound state
\be
\psi_b (y) = \sqrt {2|\eta|}\, \e^{\eta y} \, ,\qquad y\in \RR_+\, .
\label{bs}
\end{equation}
The complete orthonormal set of eigenfunctions reads now
$\{\psi(\lambda,y),\, \psi_b (y)\}$.
As a consequence, the KK measure $\varrho $ acquires \cite{Mintchev:2001mf} a
$\delta$-like contribution, localized at $M^2-\eta^2$. The theory thus
satisfies the spectral condition for $-M\leq \eta < 0$ and has a tachyon brane
excitation for $\eta < -M$.

{}For definiteness we have considered in this paper
$s=1$ noncompact extra dimensions. Both the framework and the 
results, however, admit
a direct generalization to $s\geq 1$. In that case $\lambda \mapsto 
(\lambda_1,...,\lambda_s)$,
where $\lambda_i$ has discrete spectrum if the associated dimension is compact.

Summarizing, we have demonstrated that the field
dynamics on the brane is deeply influenced by the initial
conditions in the bulk. The latter determine
the KK measure, which is a fundamental ingredient
of any theory with extra dimensions. Our analysis
reveals also the existence of a large variety of quantum fields,
generated by the same local action in the bulk, but corresponding
to different initial conditions. These fields fall into three 
classes, having distinct
locality properties. All of them preserve positivity, Poincar\'e invariance
and the spectral condition (for $\eta > -M$) on ${\bf M}_{3+1}$. It 
turns out that gradually
relaxing local commutativity, first in the bulk and after that on the 
brane, allows to derive
renormalizable, and even finite, theories.
Extra dimensions therefore emerge as a promising tool for 
constructing fundamental,
and not only effective, 3+1 dimensional quantum field theories.
In this respect, the idea of implementing a string-like
nonlocality \cite{Eliezer:1989cr} on the brane, by imposing suitable 
nonlocal initial conditions
in the bulk, looks very attractive. It will be interesting to extend 
our results to more realistic
models, which involve gauge and gravitational interactions. Among 
others, one possibility
will be to impose type ${\rm B}_1$ and ${\rm B}_2$ initial conditions 
in the gauge and
gravitational sectors
respectively, aiming to induce a unified renormalizable theory on the brane.
Some variants of this scenario are currently under investigation \cite{Mp}.

\end{document}